\documentclass[twocolumn,amsmath,amssymb,aps,showkeys]{revtex4-1}
\usepackage{graphicx}
\usepackage{amsmath,amssymb}
\usepackage{mathrsfs}
\usepackage{color}
\usepackage{afterpage}
\usepackage[version=3]{mhchem}
\usepackage{natbib}
\usepackage{soul}
\usepackage[caption=false]{subfig}
\usepackage{array}
\usepackage{multirow}
\usepackage{float}
\usepackage{braket}
\usepackage{amsmath}
\usepackage{bm}

\begin{document}

\title{Rashba dominated spin-splitting in the bulk ferroelectric oxide perovskite KIO$_3$}

\author{Sajjan Sheoran\footnote{sajjan@physics.iitd.ac.in}, Manish Kumar, Preeti Bhumla, Saswata Bhattacharya\footnote{saswata@physics.iitd.ac.in}} 
\affiliation{Department of Physics, Indian Institute of Technology Delhi, New Delhi 110016, India}

\begin{abstract}
	\noindent 
	The momentum-dependent Rashba and Dresselhaus spin-splitting has gained much attention for its highly promising applications in spintronics. In the present work, \textit{ab initio} density functional theory calculations are performed to study the spin-splitting effect in ferroelectric oxide perovskite KIO$_3$. %Non-centrosymmetric structure and presence of heavy element I, lead to momentum-dependent spin-splitting of bands. 
	Our calculations are additionally supported by symmetry adapted two-band {\textbf{\textit{k.p}}} Hamiltonian. Non-negligible spin-splitting effect is observed at conduction band minimum (CBm) and valence band maximum (VBM) for rhombohedral $R3m$ and $R3c$ phases. Linear Rashba terms successfully explain the splitting at VBM. However, cubic terms become important in realizing spin-orientation near CBm. Our results show the enhancement in Rashba parameters on tuning the ferroelectric order parameter. Further, we have observed reversal of spin-orientation on switching the direction of polarization.
\end{abstract}
\maketitle

\section{Introduction}
%In recent years, discovery of  has widely encouraged research in condensed matter physics.
Ferroelectric Rashba semiconductors have recently created a huge sensation in the field of spintronics owing to their robust spontaneous electrical polarization~\cite{10.3389/fphy.2014.00010,martin2016thin,C9NR10865C,doi:10.1021/acs.nanolett.0c03161}. These materials find applications in spin field effect transistors, ferroelectric tunnel junctions, storage and memory devices
~\cite{garcia2014ferroelectric,10.3389/fphy.2014.00010,wang2020spin}. The long-range order dipoles aligned in same direction induce polarization in ferroelectric materials, leading to inversion asymmetry. %FE polarization can be modified with the application of axial strain. 
Ferroelectric Rashba semiconductors interlink the phenomena of Rashba-type splitting and ferroelectricity, enabling the electric control of electrons' spin. In ferroelectric Rashba semiconductors, Rashba parameters can be enhanced with the application of ferroelectric (FE) polarization. Interestingly, the spin-orientation can be inverted by reversing the direction of polarization using external electric field~\cite{da2016rashba}. Electrical control of spin degree of freedom makes them suitable for spintronic devices. Some well known examples of ferroelectric Rashba semiconductors are KTaO$_3$~\cite{king2012subband}, BiAlO$_3$~\cite{da2016rashba}, LiZnSb~\cite{narayan2015class} and FASnI$_3$ (FA=Formamidinium)~\cite{stroppa2014tunable}. 

Ferroelectric materials, because of their robust spontaneous electrical polarization, are widely used in various applications. GeTe was the first ferroelectric material, in which Rashba effect was predicted theoretically~\cite{di2013electric} and observed experimentally~\cite{liebmann2016giant}. However, it has a tendency to form Ge vacancy~\cite{doi:10.1063/1.1539926}, which leads to $p$-type semiconducting behavior. This in turn poses a challenge to electric control of the spin. %The other drawbacks, including stability of halide perovskites and presence of toxic element Pb, make search of new FE materials inevitable.
%In view of this, here we have chosen a class of material that is known to be dynamically more stable to show promising applications in solar cell based devices~\cite{C8SE00451J,kumar2020theoretical}. 
In this context, ferroelectric oxide perovskites such as KIO$_3$ (KIO) show excellent piezoelectric, pyroelectric and non-linear optical properties~\cite{doi:10.1063/1.107469,kader2013charge}. KIO, in particular, is experimentally synthesized at high temperature in non-centrosymmetric rhombohedral phase with $R3m$ space group symmetry~\cite{kasatani2003study}. A dynamically favorable rhombohedral $R3c$ phase is also theoretically predicted ~\cite{bayarjargal2012phase,kader2013charge,crane1975symmetry}. In both phases, distortion of octahedra centered at I-atom induces spontaneous FE polarization. Furthermore, the presence of heavy element (I), contributing to significant spin-orbit coupling (SOC) and inversion asymmetric nature may induce interesting Rashba- and Dresselhaus-type band splitting.

%SOC and symmetry properties make a pivotal role in observing exciting phenomena such as quantum spin hall effect~\cite{PhysRevLett.96.106802,PhysRevLett.95.226801}, topological states~\cite{hasan2010colloquium} and valley degree of freedom~\cite{pacchioni2020valleytronics} in condensed matter physics.
Note that SOC and broken inversion symmetry play a pivotal role for the materials to exhibit Rashba and Dresselhaus effect.
In crystals, lacking inversion symmetry, a relativistically moving electron experiences a Lorentz-transformed magnetic field due to a finite potential gradient. This results in spin-based splitting of degenerate bands at non-time-reversal-invariant \textit{k}-points, which lifts the Kramer’s degeneracy leading to Rashba and Dresselhaus splitting. The spin-orientation is determined by the momentum dependent spin-orbit field. For acentric non-polar crystals, Dresselhaus was the first to show band splitting, which has  a cubic dependence on momentum for zincblende-type crystal structures~\cite{PhysRev.100.580}. %containing \textit{T}$_{d^2}$ point group symmetry.
For gyrotropic point group symmetries, linear Dresselhaus-type spin-splitting can also be realized. 
In polar crystals and 2D electron gas, linear splitting terms are allowed as shown by Rashba and Bychkov~\cite{Rashba1960properties,1979ZhPmR..30..574V,Bychkov_1984}.
The SOC Hamiltonian $H_{SO}=\bm{\Omega}(\textbf{\textit{k}}).\bm{\sigma}$ describes these effects,
where $\bm{\sigma}$ is pauli matrices vector and $\bm{\Omega}(\textbf{\textit{k}} )$ is spin-orbit field. The latter is odd in momentum space (i.e. $\bm{\Omega}(-\textbf{\textit{k}})=-\bm{\Omega}(\textbf{\textit{k}} )$) to preserve the time-reversal symmetry of $H_{SO}$. $\bm{\Omega}(\textbf{\textit{k}})$ depends on the spatial symmetry of the system. For simplest case, C$_{2v}$ point group symmetry, $\bm{\Omega}(\textbf{\textit{k}})$ can be written as vector sum of linear Rashba ($\bm{\Omega}_R=\alpha_R(k_y,-k_x,0)$) and Dresselhaus ($\bm{\Omega}_D=\alpha_D(k_y,k_x,0)$) spin-orbit fields~\cite{tao2021perspectives}. Here, $\alpha_R$ and $\alpha_D$ are the Rashba and Dresselhaus coefficients, respectively. These coefficients mainly depend on the amount of SOC and symmetry of the crystal~\cite{tao2017reversible}. Rashba and Dresselhaus effects lead to the same type of band splitting. However, type of splitting can be characterized by projecting spin-orientation in Fourier space, usually referred as spin texture~\cite{winkler2003spin}.

%Structural confinement is known to break the surface inversion symmetry, leading to Rashba effect, which is mainly observed in surfaces, interfaces, 2D materials  and heterostructures~\cite{liu2013tunable,mak2021polarization,lashell1996spin,koroteev2004strong,nakamura2012experimental,king2012subband,caviglia2010tunable,chakraborty2020perovskite,doi:10.1016/S1468-6996(03)00006-8}. A few examples of these systems are Au(111)~\cite{lashell1996spin}, Bi(111)~\cite{koroteev2004strong},
% due to presence of large spin-orbit interaction. 
%Some Rashba sensitive surfaces of oxides are
%SrTiO$_3$(001)~\cite{nakamura2012experimental}, KTaO$_3$(001)~\cite{king2012subband}, LaAlO$_3$/SrTiO$_3$~\cite{caviglia2010tunable}, InP/InGaAs~\cite{doi:10.1016/S1468-6996(03)00006-8} and LaAlO$_3$/SrIrO$_3$~\cite{chakraborty2020perovskite}. Non-centrosymmetric bulk materials with significant amount of SOC such as BiTeX (X=I, Br, Cl)~\cite{ishizaka2011giant}, BiAlO$_3$~\cite{da2016rashba}, LaWN$_3$~\cite{bandyopadhyay2020origin} and CH$_3$NH$_3$PbI$_3$~\cite{frohna2018inversion} have also been reported as Rashba-sensitive. Contrarily, Dresselhaus effect is mainly observed in zincblende-type bulk crystal structures. Bulk materials viz. HfO$_2$~\cite{tao2017reversible}, FASnI$_3$ (FA=Formamidinium)~\cite{stroppa2014tunable} and BiAlO$_3$(BAO)~\cite{da2016rashba} show both Rashba and Dresselhaus effects.

In this article, we have studied the Rashba and Dresselhaus effects in $R3m$ and $R3c$ phases of FE oxide KIO$_3$ using state-of-the-art density functional theory (DFT) and symmetry adapted two-band \textbf{\textbf{\textit{k.p}}} Hamiltonian. Firstly, we have determined the FE polarization in both the phases. Subsequently, the electronic atom-projected partial density of states (pDOS) and band structures are determined using DFT. The crucial effect of SOC has been shown in the band structures. The Rashba spin-splitting energy and offset momentum have been determined from the splitting at valence band maximum (VBM) and conduction band minimum (CBm). Further, the type of splitting has been characterized by plotting the spin texture. The Rashba and Dresselhaus parameters are determined after fitting the \textbf{\textbf{\textit{k.p}}} Hamiltonian to the DFT band structure. Finally, the effect of polarization on the aforementioned parameters has been investigated.  %The work is organized as follows: in Sec. II, we have discussed the computational methods used in the simulations. Further, we have discussed structural details and calculated the FE properties of KIO in Sec. III. Sec. IV is completely devoted to detailed electronic properties  and \textbf{\textbf{\textit{k.p}}} model analysis. Finally, conclusions are drawn in Sec. V.

\section{COMPUTATIONAL METHODS}
The calculations are performed using Vienna \textit{ab initio} simulation package (VASP)~\cite{kresse1993ab,kresse1996efficient} within the framework of DFT using projector augmented wave (PAW) pseudopotentials. The Perdew-Burke-Ernzerhof (PBE) exchange-correlation ($\epsilon_{xc}$) functional is used for DFT calculations~\cite{perdew1996generalized}. For better accuracy of excited state properties and validations of PBE results, non local Heyd-Scuseria-Ernzerhof (HSE06) $\epsilon_{xc}$ functional is used~\cite{doi:10.1063/1.1564060}. For effective interpretation of results, conventional hexagonal setting is also considered. A cutoff energy of 600 eV is used throughout the calculations. Rhombohedral phases $R3c$  and $R3m$ are relaxed without including SOC with  9$\times$9$\times$4 and 9$\times$9$\times$8 \textit{k}-grids, respectively, generated using Monkhorst-Pack scheme~\cite{pack1977special}. Insignificant role of SOC in relaxation is also verified by test calculations. The pDOS and band structure calculations are done using 12$\times$12$\times$6 \textit{k}-grid. Spin texture calculations are done with  closely spaced 13$\times$13 \textit{k}-grid around high symmetry points. In structural optimization, the total energy difference between two ionic  relaxation steps is set to smaller than $10^{-5}$ eV and tolerance on forces between two consecutive steps is set to 0.001 eV/\AA. FE properties are calculated within the framework of berry phase theory for polarization~\cite{king1993theory,resta1994macroscopic,spaldin2012beginner}. %Note that the SOC is also included in the single point energy calculations. 
Spin textures are calculated using expectation values of spin operators $S_{i}$ (s$_i$=$\langle S_i \rangle$) {\footnote{$s_i=\bra{\Psi_k} {S_i} \ket{\Psi_k}$=$\frac{\hbar}{2} \bra{ \Psi_k} \sigma_i \ket {\Psi_k}$, $\hbar$=1.}} (\textit{i}=\textit{x}, \textit{y}, \textit{z}), given by
\begin{equation}
s_i=\frac{1}{2} \bra{ \Psi_k} \sigma_i \ket {\Psi_k}
\end{equation}
where $\sigma_i$ are the pauli matrices and $\Psi_k$ is the spinor eigenfunction obtained from noncollinear spin calculations. 
\begin{table}
    \begin{center}
            \caption {Lattice parameters and polarization for rhombohedral phases of KIO.}
            \label{T1}
            \begin{tabular}{ |c|c|c|c|c|}
            	\hline
            	Space group & a(\AA) & c(\AA) & V(\AA $^3$) & P($\mu$C/cm$^2$) \\ \hline
            	$R3c$         &  6.37  & 15.91  & 558.6       &        29         \\ \hline
            	$R3m$         &  6.29  &  8.11  & 278.1       &        41         \\ \hline
            \end{tabular}
   \end{center}
\end{table}

\section{STRUCTURAL AND FERROELECTRIC PROPERTIES}
%KIO shows rich phase transitions with structures having \textit{P}1, \textit{Pm}$\bar{3}$\textit{m}, \textit{R}3, $R3m$ and $R3c$ space group symmetries.
KIO mainly exists in \textit{Pm}$\bar{3}$\textit{m}, $R3m$ and $R3c$ space group symmetries. \textit{Pm}$\bar{3}$\textit{m} phase is centrosymmetric i.e., contains an inversion center and therefore, does not show the Rashba-type splitting. Hence, we have studied the non-centrosymmetric symmorphic rhombohedral phases $R3m$ and $R3c$ in detail for Rashba and Dresselhaus properties. 
The structural details used in the present work are provided in Table~\ref{T1}. 
For $R3m$ phase, we have calculated  change in ferroelectric polarization i.e., dipole moment per unit volume ($\frac{\textbf{p}}{V}$) with respect to centrosymmetric structure, of 41 $\mu$C/cm$^2$  along the [0001] direction in hexagonal setting (along the [111] direction in rhombohedral setting). However, in $R3c$ phase, we have found slightly smaller polarization of 29 $\mu$C/cm$^2$ along the [0001] direction in hexagonal setting. % The direction of polarization is same as in BAO with somewhat smaller magnitude.
%Therefore, $R3m$ phase is more FE than $R3c$ phase by 12 $\mu$C/cm$^2$.
In previous studies on BiFeO$_3$ thin films, it has been shown that FE polarization can be enhanced to as large as 150 $\mu$C/cm$^2$ using the strain ~\cite{PhysRevLett.95.257601,Ricinschi_2006,doi:10.1063/1.4863778}. In view of this, we have also verified the enhancement in FE polarization of KIO on the application of strain. The FE polarization has increased from 29 to 35 $\mu$C/cm$^2$ and 41 to 50 $\mu$C/cm$^2$ in $R3m$ and $R3c$ phases, respectively, on applying uniaxial strain of 10\% in the direction of polarization (in \textit{z}-direction) within the harmonic approximation (see discussion later). 

\section{ELECTRONIC PROPERTIES AND Rashba-Dresselhaus EFFECTS}
\begin{figure}[htp]
	\centering
	\includegraphics[width=8cm]{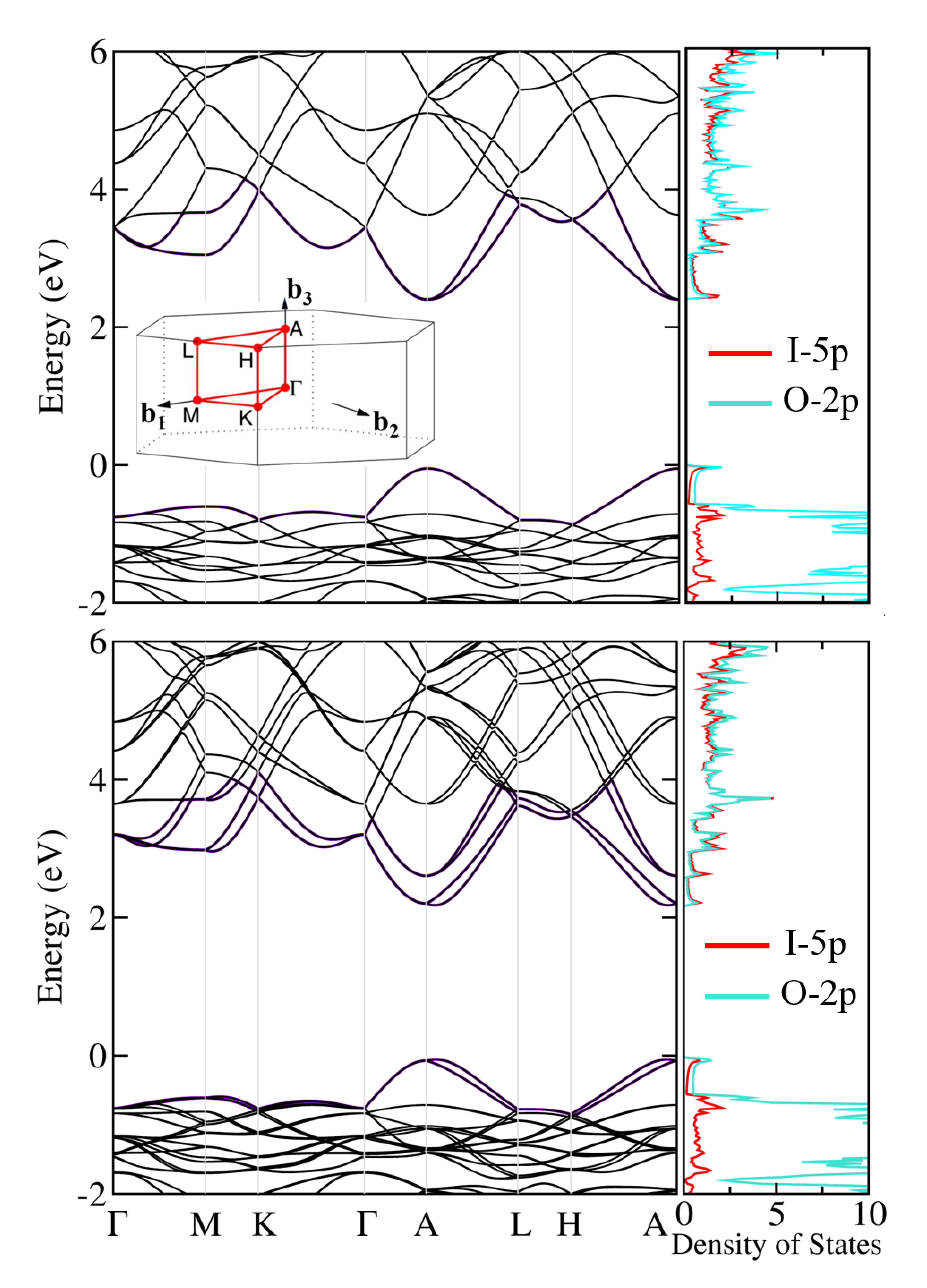}
	\caption{Band structure and pDOS for $R3m$ phase calculated using PBE (a) without SOC and (b) with SOC. The Fermi energy is set to VBM. Here, inset shows the first Brillouin zone for the hexagonal phase.}
	\label{fig_1}
\end{figure}
Fig.~\ref{fig_1}a shows the calculated band structure for the $R3m$ phase without SOC along the high-symmetry path (HSP) in the first Brillouin zone (see Fig.~\ref{fig_1}a inset for HSP). A direct band gap of 2.51 eV is observed.  Since PBE is known to underestimate the band gap, it is also calculated using HSE06. A larger direct band gap of 3.50 eV is observed at the \textit{k}-point A with no significant changes in band profile. The CBm and VBM occur at the \textit{k}-point A. The uppermost valence band has a width of about 2.5 eV and the electronic states are mainly derived from O-2p orbitals (see pDOS in Fig.~\ref{fig_1}). The lowest conduction band has a width of nearly 4 eV and the electronic states are mainly derived from equal contribution of I-5p and O-2p orbitals. Fig.~\ref{fig_1}b shows the calculated band structure and pDOS with inclusion of SOC. The VBM and CBm shift from \textit{k}-point A towards L, which is known as the offset momentum ($\delta k$).  A slightly indirect band gap of 2.24 and 3.27 eV is observed using PBE+SOC and HSE06+SOC, respectively. Despite the underestimation of the band gap by PBE, the band dispersion around the high symmetry point is known to be similar to HSE06~\cite{di2013electric}. We have compared the band structures obtained using PBE+SOC and HSE06+SOC and found that they are resulting in similar Rashba parameters (see Section I of Supplemental Information (SI)). Therefore, all the calculations are performed using PBE, since it is more cost effective. The presence of large SOC is attributed to the heavy elements like I. Dominant spin-splitting can be seen in the plane \textit{k}$_z$=$\frac{\pi}{c}$, which is perpendicular to the polarization axis (see Fig.~\ref{fig_1}b along A-L and A-H directions). In contrast, splitting is completely absent in direction $\Gamma$-A, which is parallel to the polarization axis [0001]. It is consistent with the Rashba model, where splitting occurs in direction perpendicular to the polarization axis. %A predominant Rashba-type splitting can be seen near VBM and CBm in directions A-L and A-H (see Fig. 2). 
The energy difference between the \textit{k}-point A and extremum is known as the Rashba spin-splitting energy ($\delta E$).

\begin{figure}[htp]
	\centering
	\includegraphics[width=8.5cm]{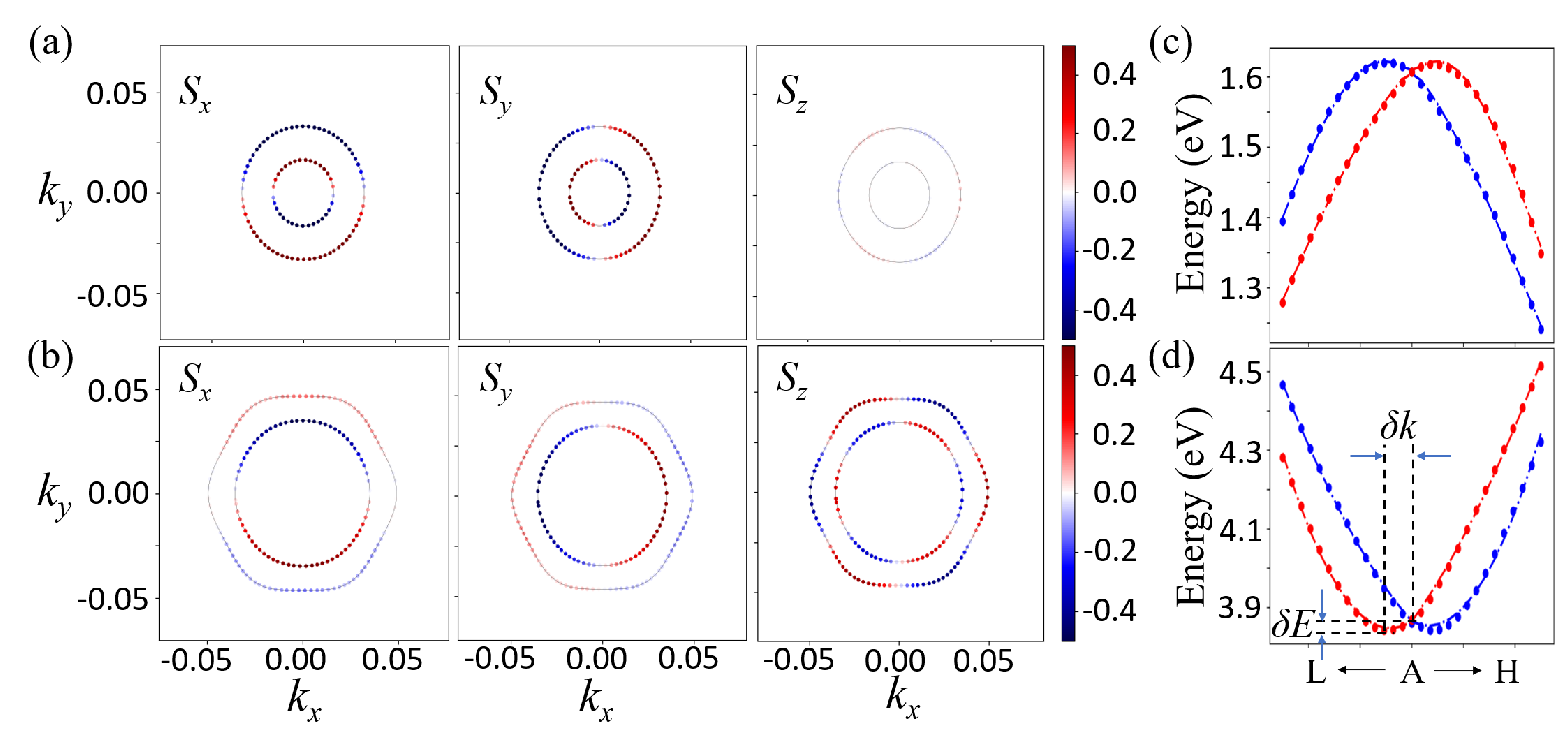}
	\caption{Spin texture in $R3m$ phase, calculated at constant energy surface, near (a) VBM (E=E$_\textrm{F}$$-$0.2 eV) and (b) CBm (E=E$_\textrm{F}$$+$2.5 eV) centered at \textit{k}-point A. Spin-splitting of the (c) VBM and (d) CBm around \textit{k}-point A. Band structure is plotted along direction ($\frac{2\pi}{a}$0.25,0,0.5)-(0,0,0.5)-($\frac{2\pi}{a}$0.16,$\frac{2\pi}{b}$0.16,0.5) of Fourier space, which is along L-A-H. DFT and \textbf{\textit{k.p}} band structures are plotted with dashed lines and dots, respectively. Here, the color is representing the spin-projection.}
	\label{fig_2}
\end{figure}

In order to have better understanding of the spin-splitting nature, spin texture is plotted near the VBM and CBm around \textit{k}-point A. The 2D spin texture is calculated by projecting expectation values of $\sigma_x$, $\sigma_y$ and $\sigma_z$ in Fourier plane (\textit{k}$_x$-\textit{k}$_y$) using PyProcar~\cite{herath2020pyprocar}. Fig.~\ref{fig_2}a and~\ref{fig_2}b show the calculated \textit{x}-, \textit{y}- and \textit{z}-component of spin texture near VBM and CBm, respectively. The in-plane spin components (\textit{S$_x$}, \textit{S$_y$}) show helical nature with inner and outer bands having opposite orientation. This confirms the existence of Rashba-type splitting. A significant out of plane spin component can also be seen near CBm, which is absent near VBM. The out of plane spin component ($S_z$)  has three-fold symmetry, which is in agreement with the three fold rotation symmetry of the crystal. The little group of \textit{k}-point A  is $C_{3v}$, consisting of three-fold rotations $C_3$, one reflection through vertical plane containing \textit{z}-axis ($\sigma_{xz}$) and two reflections through diagonal planes ($M_{d1}$, $M_{d2}$)~\cite{link1,link2} (besides trivial identity operation). Band dispersion relation and spin texture around \textit{k}-point A in plane orthogonal to polar axis can be derived using all the symmetry-allowed terms such that $O^{\dagger}H(\textbf{\textit{k}})O=H(\textbf{\textit{k}})$, where $O$ is the symmetry operation belonging to the little group~\cite{da2016rashba}. The constructed two-band \textbf{\textbf{\textit{k.p}}} Hamiltonian including linear and cubic Rashba terms satisfying the $C_{3v}$ symmetry near \textit{k}-point A takes the form~\cite{vajna2012higher} (for more details see Section II in SI)   
\begin{equation}
  H_A(\textbf{\textit{k}})=H_o(\textbf{\textit{k}})+ H_{SO}
  \label{Eq_1}
\end{equation}
where, 
\begin{equation}
H_{SO}=\alpha \sigma_y k_x + \beta \sigma_x k_y 
+ \gamma \sigma_z[ (k_x^3+k_y^3)-3(k_xk_y^2+k_yk_x^2)]
\end{equation}
 and $H_o(\textbf{\textit{k}})$ is free particle Hamiltonian. $\alpha$, $\beta$ are the coefficients  of linear terms and $\gamma$ is the coefficient of cubic term in SOC Hamiltonian. Two energy eigenvalues of Hamiltonian are\\
\begin{equation}
  E(\textbf{\textit{k}})^{\pm}= \frac{\hbar^2 k_x^2}{2m_x}+\frac{\hbar^2 k_y^2}{2m_y} \pm E_{SO}  
\end{equation}
where, $m_x$ and $m_y$ represent the effective masses in $x$ and $y$ directions, respectively. $E_{SO}$ is the energy eigenvalue of SOC Hamiltonian given by $E_{SO}(\textbf{\textit{k}})=\sqrt{\alpha^2 k_x^2+\beta^2 k_y^2+ \gamma^2 f^2(k_x,k_y)}$, where $f(k_x,k_y)=(k_x^3+k_y^3)-3(k_xk_y^2+k_yk_x^2)$. Normalized spinor wavefunctions corresponding to energy eigenvalues are given by 
\begin{equation}
\Psi_{\textbf{\textit{k}}}^{\pm} = \frac{e^{i\textbf{\textit{k.r}}}}{\sqrt{2\pi(\rho^2_{\pm}+1 })}\begin{pmatrix}
\frac{i\alpha k_x-\beta k_y}{\gamma f(k_x,k_y)\mp E_{SO}} \\
1
\end{pmatrix}
\end{equation}
where $\rho^2_{\pm}=\frac{\alpha^2 k_x^2+\beta^2 k_y^2}{(\gamma f(k_x,k_y)\mp E_{SO})^2}$. The expectation values of spin operators are given by 
\begin{equation}
\{s_x,s_y,s_z\}^{\pm}=\pm \frac{1}{E_{so}}\{\beta k_y,\alpha k_x,\gamma f(k_x,k_y) \}
\end{equation}
Spin orientation in \textit{x} (\textit{y}) direction depends on $k_y$ ($k_x$) and it becomes zero at \textit{k}$_y$=0 (\textit{k}$_x$=0). In-plane spin components are reproduced using $\alpha$ and $\beta$, whereas $\gamma$ reproduces out of plane spin component. In-plane spin components are small if $\gamma$ is much larger than $\alpha$ and $\beta$, and vice versa. Spin texture calculated using DFT satisfies the model Hamiltonian predictions. Three-fold degeneracy of \textit{z}-component of spin is the consequence of cubic nature of $f(k_x,k_y)$ (see Section II in SI for more details).\\ 
   
  Fig.~\ref{fig_2}c and~\ref{fig_2}d show the comparison between the DFT and \textbf{\textbf{\textit{k.p}}} model predicted band structures near VBM and CBm, respectively, in the vicinity of \textit{k}-point A. The \textbf{\textit{k.p}} model produces band structure, which is in close agreement with the DFT band structure. Near the \textit{k}-point A, cubic terms have negligible contribution in band structure calculations that allows to estimate the values of $\alpha$ and $\beta$. Hence, energy eigenvalues of the Hamiltonian are given by $E(\textbf{\textit{k}})^{\pm}=\frac{\hbar^2 k_x^2}{2m_x}+\frac{\hbar^2 k_y^2}{2m_y}  \pm \sqrt{\alpha^2k_x^2+\beta^2k_y^2}$, which estimate only the magnitude of $\alpha$ and $\beta$. The signs of $\alpha$ and $\beta$ are determined by the orientation of spins in Fourier space. Rashba and Dresselhaus coefficients are defined as $\alpha_R=\frac{\alpha-\beta}{2}$ and $\alpha_D=\frac{\alpha+\beta}{2}$, respectively (the details can be seen in Section II of SI). For VBM, $\delta E$=15.1 meV and $\delta k$=0.062 \AA$^{-1}$ are obtained from the DFT band structure along HSP L-A (see Fig.~\ref{fig_2}c). It provides $\alpha$=$2\delta E/\delta k=0.49$ eV\AA.  %For VBM, from the DFT calculations in L-A direction, calculated $\delta$= 15.1 meV and $\delta$k= 0.062\AA$^{-1}$ gives $\alpha$= $2 \delta E/\delta k=0.49$ eV\AA.
   In A-H direction, Rashba spin-splitting is same as in L-A direction. However, the value of $\delta k$ is 0.051 \AA$^{-1}$ (see Fig.~\ref{fig_2}c). Thus, fitting the DFT band structure in that direction gives 
   $\sqrt{\alpha^2+\beta^2}$=$2 \delta E/\delta k$=0.59 eV\AA. It results into $\beta$=$-$0.33 eV\AA. $\alpha_R$ and $\alpha_D$ are found to be 0.41 eV\AA\, and 0.08 eV\AA\,, respectively, using  $\alpha$ and $\beta$. For CBm, in L-A and A-H directions, a larger Rashba spin-splitting of 23.2 meV is calculated (see Fig. ~\ref{fig_2}d). The offset momentum is observed to be 0.054 \AA$^{-1}$ and 0.042 \AA$^{-1}$ in L-A and A-H directions, respectively (see Fig. ~\ref{fig_2}d). Using the same approach as for VBM, $\alpha_R$=0.77 eV\AA\, and $\alpha_D$=0.11 eV\AA\, are calculated. All the calculations are summarized in Table~\ref{T2}. We have observed comparatively larger Rashba splitting at CBm than VBM due to the higher contribution of I-5p orbitals at CBm (see Fig. \ref{fig_1}). The model Hamiltonian with $\alpha$=0.50 eV\AA, $\beta$=$-$0.33 eV\AA\, and $\gamma$=$-$0.06 eV\AA$^3$ reproduces the band structure and spin texture near \textit{k}-point A for VBM, that are well in agreement with the DFT predictions (see Fig. \ref{fig_2}a and \ref{fig_2}c). Similarly, for CBm, $\alpha$=0.87 eV\AA, $\beta$=$-$0.67 eV\AA\, and $\gamma$=$-$4.38 eV\AA$^3$ reproduce the DFT band structure and spin texture (see Fig. \ref{fig_2}b and \ref{fig_2}d).
\begin{table}
 	\begin{center}
 		\caption {Rashba parameters for band-splitting at \textit{k}-point A for $R3m$ phase.}
 		\label{T2}
 		\begin{tabular}{|p{1.30cm}|p{1.25cm}|p{1.25cm}|p{1.25cm}|p{1.20cm}|p{1.20cm}|}
 			\hline
 			Position& $\delta E$ & $\delta$\textit{k}$_{\textrm{A-L}}$ & $\delta$\textit{k}$_\textrm{A-H}$ & $\alpha_R$& $\alpha_D$\\
 			& (meV) &(\AA$^{-1}$)& (\AA$^{-1}$) & (eV\AA) & (eV\AA)  \\ \hline
 			VBM         &  15.1  & 0.062 & 0.051& 0.41     &        0.08       \\ \hline
 			CBm        &  23.2  &  0.054  & 0.042 & 0.77     &        0.11        \\ \hline
 		\end{tabular}
 	\end{center}
 \end{table}    
\begin{figure}[htp]
	\centering
	\includegraphics[width=7cm]{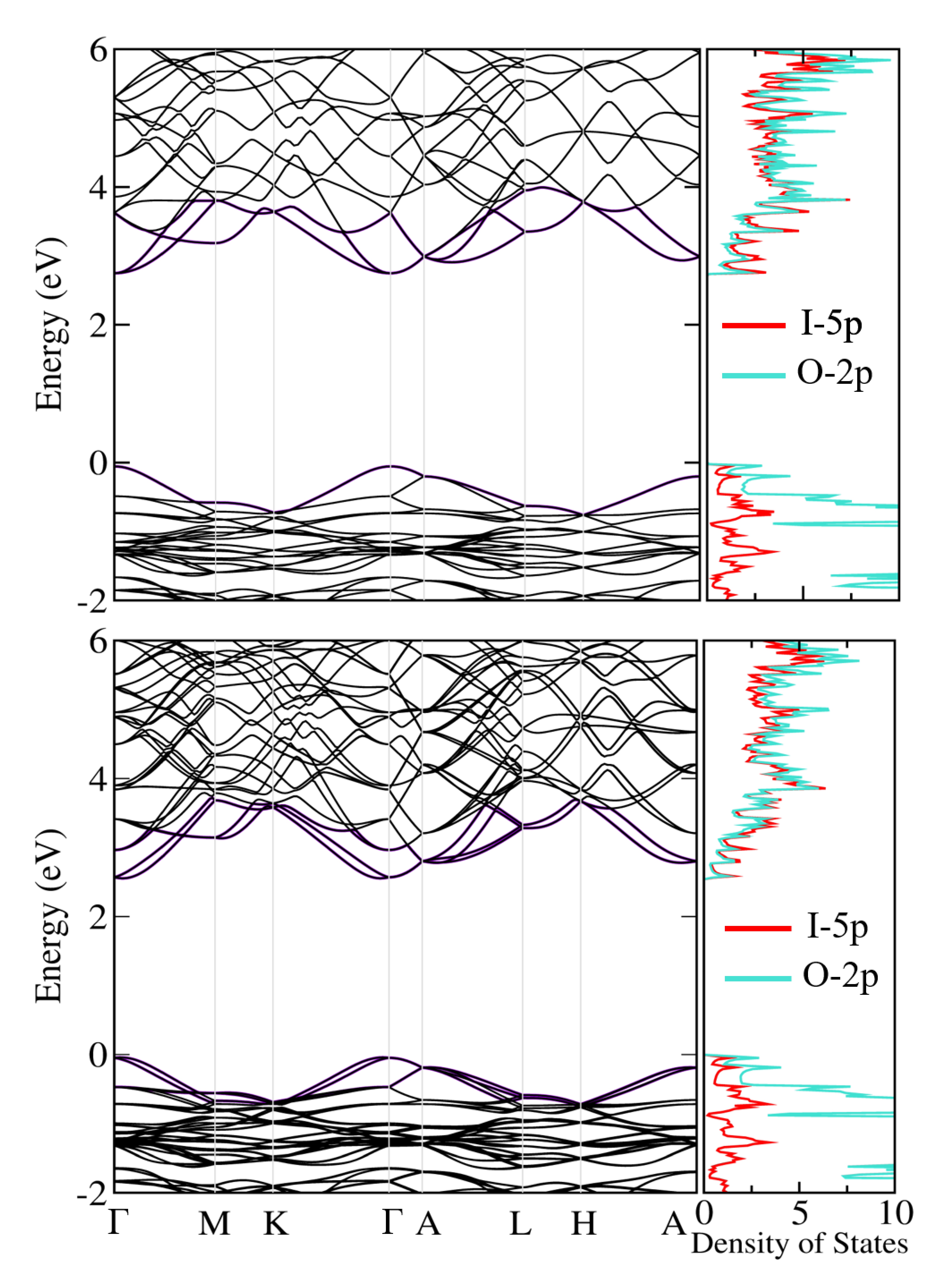}
	\caption{Band structure and pDOS for $R3c$ phase calculated using PBE (a) without SOC and (b) with SOC. The Fermi energy is set to VBM.}
	\label{fig_3}
\end{figure}

\begin{figure}[htp]
	\centering
	\includegraphics[width=8.5cm]{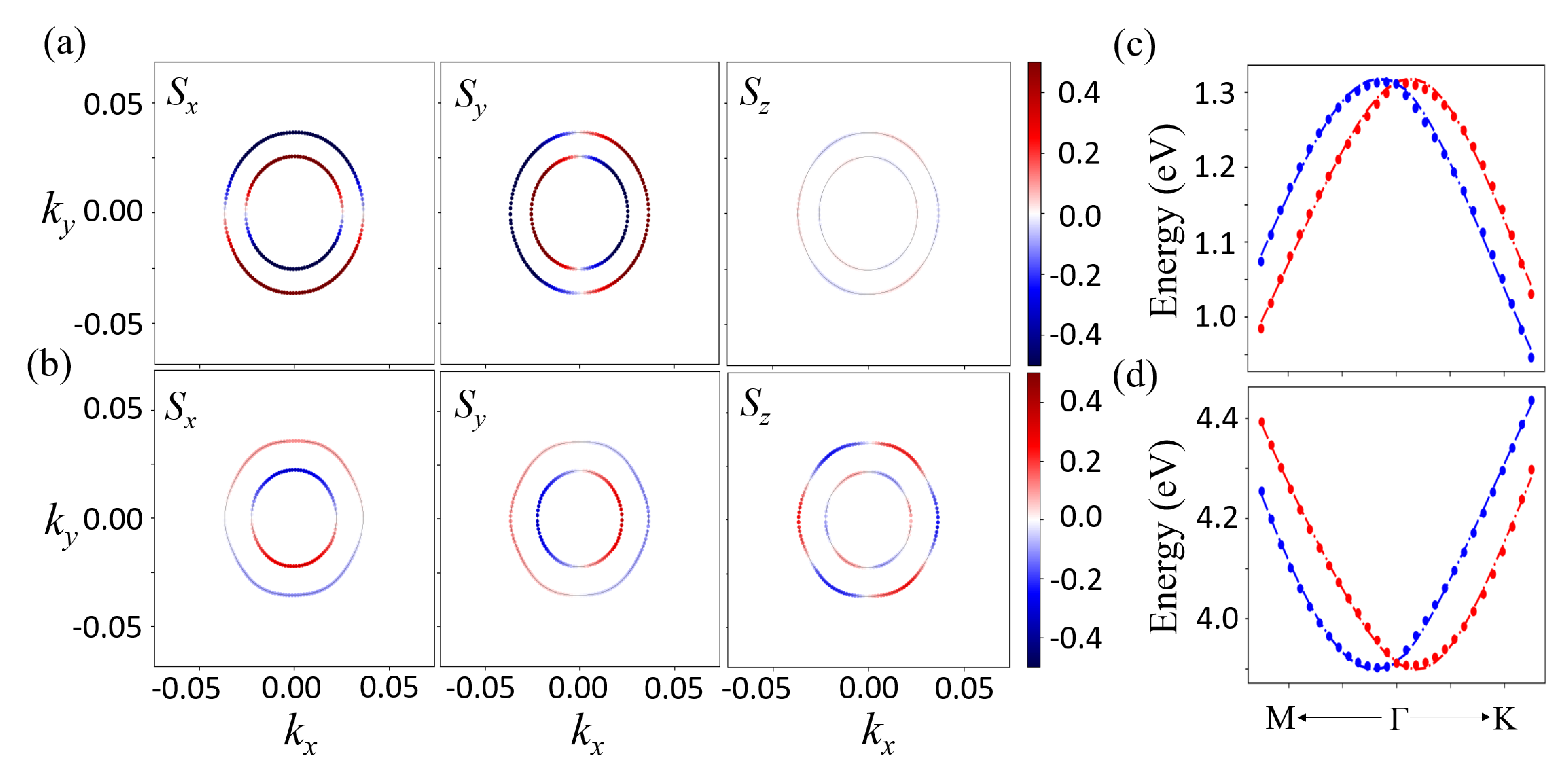}
	\caption{Spin texture in $R3c$ phase, calculated at constant energy surface, near (a) VBM (E=E$_\textrm{F}$$-$0.2 eV) and (b) CBm (E=E$_\textrm{F}$$+$3.0 eV) centered at \textit{k}-point $\Gamma$. Spin-splitting of the (c) VBM and (d) CBm around \textit{k}-point $\Gamma$. Band structure is plotted along direction ($\frac{2\pi}{a}$0.25,0,0)-(0,0,0)-($\frac{2\pi}{a}$0.16,$\frac{2\pi}{b}$0.16,0) of Fourier space, which is along M-$\Gamma$-K. DFT and \textbf{\textit{k.p}} band structures are plotted with dashed lines and dots, respectively. Here, the color is representing the spin-projection.}
	\label{fig_4}
\end{figure}

Fig.~\ref{fig_3}a and~\ref{fig_3}b show the band structure and pDOS for $R3c$ phase without and with inclusion of SOC, respectively. It has a direct band gap of 2.86 eV  at \textit{k}-point $\Gamma$ without including SOC, which is 0.41 eV larger than for the $R3m$ phase. The contribution of atomic orbitals in the pDOS near VBM and CBm for the $R3c$ phase are nearly similar to the $R3m$ phase. With the inclusion of SOC, VBM and CBm shift towards \textit{k}-point K. A slightly indirect band gap of 2.65 eV on including SOC confirms the importance of SOC in the calculations. A better estimate of band gap using HSE06+SOC is calculated to be 3.80 eV. Spin-splitting can be seen throughout the Brillouin zone except for $\Gamma$-A,  which again confirms the polarization direction parallel to $\Gamma$-A (see Fig.~\ref{fig_3}b). Dominant spin-splitting can be seen along HSP $\Gamma$-M and $\Gamma$-K. Fig.~\ref{fig_4}a and~\ref{fig_4}b show the plotted spin texture near VBM and CBm, respectively, around \textit{k}-point $\Gamma$. In-plane spin textures form helical-type spin texture with different orientation for inner and outer bands, confirming the Rashba-type splitting of degenerate levels. A significant out of plane spin component can also be seen near CBm, which is absent near VBM. The little group of \textit{k}-point $\Gamma$  is $C_{3v}$. The model Hamiltonian given by Eq. ~\ref{Eq_1} can explain the band properties of VBM and CBm near \textit{k}-point  $\Gamma$. Fig.~\ref{fig_4}c and~\ref{fig_4}d show that DFT and \textbf{\textit{k.p}} model predicted band structures near the \textit{k}-point $\Gamma$ are comparable. Rashba spin-splitting energy of 5.4 and 14.8 meV are obtained for VBM and CBm, respectively. The calculations for $R3c$ phase along $\Gamma$-M and $\Gamma$-K directions are done by proceeding the same as for $R3m$ phase. We have found that $\alpha_R$=0.26 eV\AA, $\alpha_D $=0.04 eV\AA\, for VBM and $\alpha_R$=0.52 eV\AA , $\alpha_D $=0.11 eV\AA\, for CBm. 
The Hamiltonian given in Eq.~\ref{Eq_1} with $\alpha$=0.30 eV\AA, $\beta$=$-$0.21 eV\AA\, and $\gamma$=$-$0.01 eV\AA$^3$ well satisfies the DFT results for splitting near VBM (see \ref{fig_4}a and \ref{fig_4}c). CBm splitting is well approximated by $\alpha$=0.63 eV\AA, $\beta$=$-$0.41 eV\AA\, and $\gamma$=$-$3.34 eV\AA$^3$ (see \ref{fig_4}b and \ref{fig_4}d). For quick review, all the parameters are summarized in Table~\ref{T3}.  Rashba coefficients of some selected ferroelectric materials are compared with KIO in Section III of SI.\\ 
\begin{table}
	\begin{center}
		\caption {Rashba parameters for band-splitting at \textit{k}-point $\Gamma$ for $R3c$ phase.}
		\label{T3}
		\begin{tabular}{|p{1.30cm}|p{1.25cm}|p{1.25cm}|p{1.25cm}|p{1.20cm}|p{1.20cm}|}
			\hline
			Position& $\delta E$ & $\delta$\textit{k}$_{\Gamma-\textrm{M}}$ & $\delta$\textit{k}$_{\Gamma-\textrm{K}}$ & $\alpha_R$& $\alpha_D$\\
			& (meV) &(\AA$^{-1}$)& (\AA$^{-1}$) & (eV\AA) & (eV\AA)  \\ \hline
			VBM         &  5.4  & 0.035 & 0.029& 0.26      &        0.04       \\ \hline
			CBm        &  14.8  &  0.047  & 0.039 & 0.52     &        0.11        \\ \hline
		\end{tabular}
	\end{center}
	
\end{table}
%\begin{figure}
%	\centering
%%	\caption{Spin-splitting of (a) CBm and (b) VBM around the \textit{k}-point $\Gamma$ in $R3c$ phase, respectively. Band structure is plotted along direction ($\frac{2\pi}{a}$0.25,0,0)-(0,0,0)-($\frac{2\pi}{a}$0.16,$\frac{2\pi}{b}$0.16,0) of Fourier space, which is M-$\Gamma$-K direction. DFT and \textbf{\textit{k.p}} band structures are plotted with dashed lines and dots, respectively. Red and green colors represent spin-up and spin-down components, respectively.}
%	\label{fig_8}
%\end{figure}
\begin{figure}[htp]
	\centering
	\includegraphics[width=6cm]{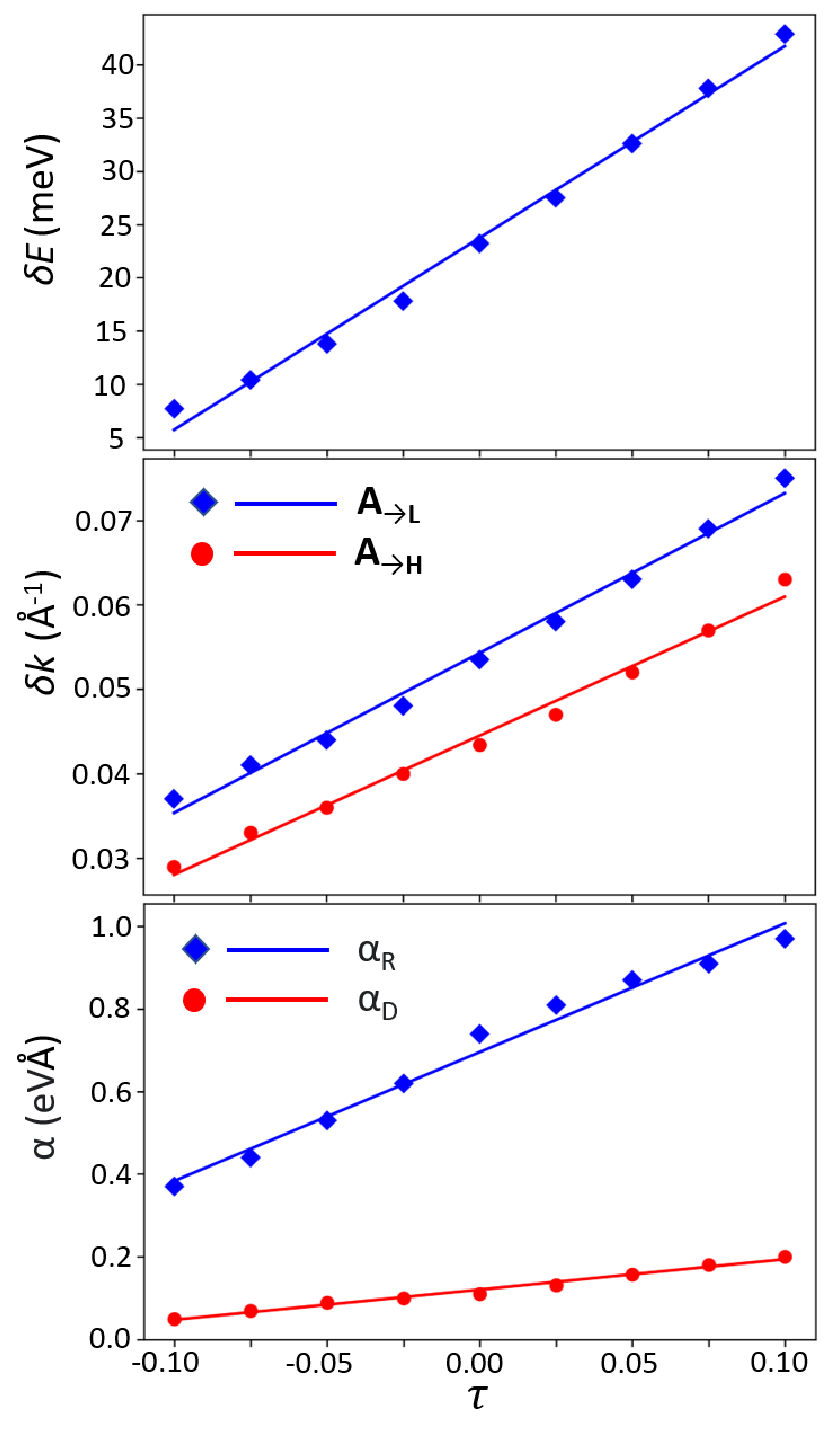}
	\caption{(a) Rashba spin-splitting energy ($\delta E$), (b) offset momentum ($\delta k$), (c) Rashba parameter ($\alpha_R$) and Dresselhaus parameter ($\alpha_D$) as a function of ferroelectric order parameter ($\tau$). The values are calculated for CBm in $R3m$ phase. The calculated values are linearly fitted. Note that $\delta E$ is same along both A-H and A-L directions.}
	\label{fig_5}
\end{figure}

Effect of ferroelectric order parameter ($\tau$) on the Rashba parameters is also studied for CBm in $R3m$ phase. $\tau$ (representing the relative distance between atoms in the unit cell) is varied from -0.10 to 0.10. Here $\tau$ is defined as 
\begin{equation}
\tau = \frac{c-c_0}{c_0}
\end{equation}

where $c$ and $c_0$ are the lattice constants of strained and unstrained unit cell, respectively. 
We have observed the linear trend in Rashba spin-splitting energy and offset momentum along HSP A-L and A-H (see Fig.~\ref{fig_5}a and~\ref{fig_5}b). The change in Rashba coefficient is  more significant as compared to Dresselhaus coefficient (see Fig.~\ref{fig_5}c). The spin degeneracy can be restored by bringing back Rashba spin-splitting and offset momentum to zero via tuning the $\tau$. The calculations on BAO showed that Rashba and Dresselhaus parameters can be made negative by switching the direction of spontaneous polarization using external electric field~\cite{da2016rashba}. The full reversal of spin texture on reversing the direction of electric field is also observed in KIO (see Fig.~\ref{fig_5}a and~\ref{fig_5}b). However, the change in parameters is significantly higher for KIO than BAO. The $\alpha_R$ can be tuned from 0.38 eV\AA\, to 1.1 eV\AA\, by varying the $\tau$ from -0.10 to 0.10 (see Fig.~\ref{fig_5}c). Rashba parameters are larger for KIO than BAO, despite significantly larger ferroelectric polarization in BAO. Therefore, we infer that larger ferroelectric polarization does not directly imply the larger splitting. A thorough analysis of symmetry and electronic structure analysis is always required. 
\section{CONCLUSION}
In summary, we have performed the relativistic first-principles density functional theory calculations to study Rashba and Dresselhaus effects in ferroelectric rhombohedral phases of KIO with $R3m$ and $R3c$ space group symmetries. Ferroelectric and electronic properties are explored using DFT, which are also supported by symmetry adapted \textbf{\textit{k.p}} Hamiltonian. Near VBM and CBm, the states are mainly derived from I-5p and O-2p orbitals. A sufficiently wide and slightly indirect band gap is calculated for $R3m$ (2.24 (PBE+SOC), 3.27 (HSE06+SOC) eV) and $R3c$  (2.65 (PBE+SOC), 3.80 (HSE06+SOC) eV) phases, respectively. Due to significant amount of SOC, spin-splitting effects are observed at both VBM and CBm around \textit{k}-points A and $\Gamma$ for $R3m$ and $R3c$ phases, respectively. Helical-type in-plane spin texture confirms that the spin-splitting mainly consists of Rashba-type splitting. Out of plane spin texture shows importance of cubic terms in model Hamiltonian. Hamiltonian satisfying C$_{3v}$ symmetry reproduces band structure and spin texture near CBm and VBM, that are well in agreement with the DFT results. The largest Rashba coefficient is found for CBm in $R3m$ phase. Full reversal of spin texture on reversing the direction of polarization is also verified. Further, we have investigated the effect of strain on Rashba and Dresselhaus parameters and found that they increase linearly with strain. Control of spin-based properties using the external electric field makes it suitable for spintronics applications. The larger Rashba coefficient than other contemporary materials (viz. BAO and LaWN$_3$)  makes KIO a promising addition into this class of material having Rashba-based applications. 
\section{Acknowledgement}
S.S. acknowledges  CSIR,  India,  for  the  junior  research fellowship [Grant No. 09/086(1432)/2019-EMR-I]. M.K.  acknowledges  CSIR,  India,  for  the  senior  research fellowship [Grant No. 09/086(1292)/2017-EMR-I]. P.B. acknowledges UGC, India, for the senior research fellow-ship [1392/(CSIR-UGC NET JUNE 2018)].  S.B. acknowledges the financial support from SERB under Core  Research  Grant  (Grant  No.  CRG/2019/000647).
We acknowledge the High Performance Computing (HPC) facility at IIT Delhi for computational resources.
\bibliography{ref}

\end{document}

\title{Supplemental Material for\\
	``Rashba dominated spin-splitting in the bulk ferroelectric oxide perovskite KIO$_3$"}
\author{Sajjan Sheoran\footnote{sajjan@physics.iitd.ac.in}, Manish Kumar, Preeti Bhumla, Saswata Bhattacharya\footnote{saswata@physics.iitd.ac.in}} 
\affiliation{Department of Physics, Indian Institute of Technology Delhi, New Delhi 110016, India}
%\date{\today}
\pacs{}
\keywords{DFT, Model Hamiltonian, Rashba, Spin-orbit coupling,\textbf{k.p} perturbation theory }
\maketitle
\section{Comparison between PBE and HSE06 band structures}

\begin{figure}[H]
	\centering
	\includegraphics[scale=0.48]{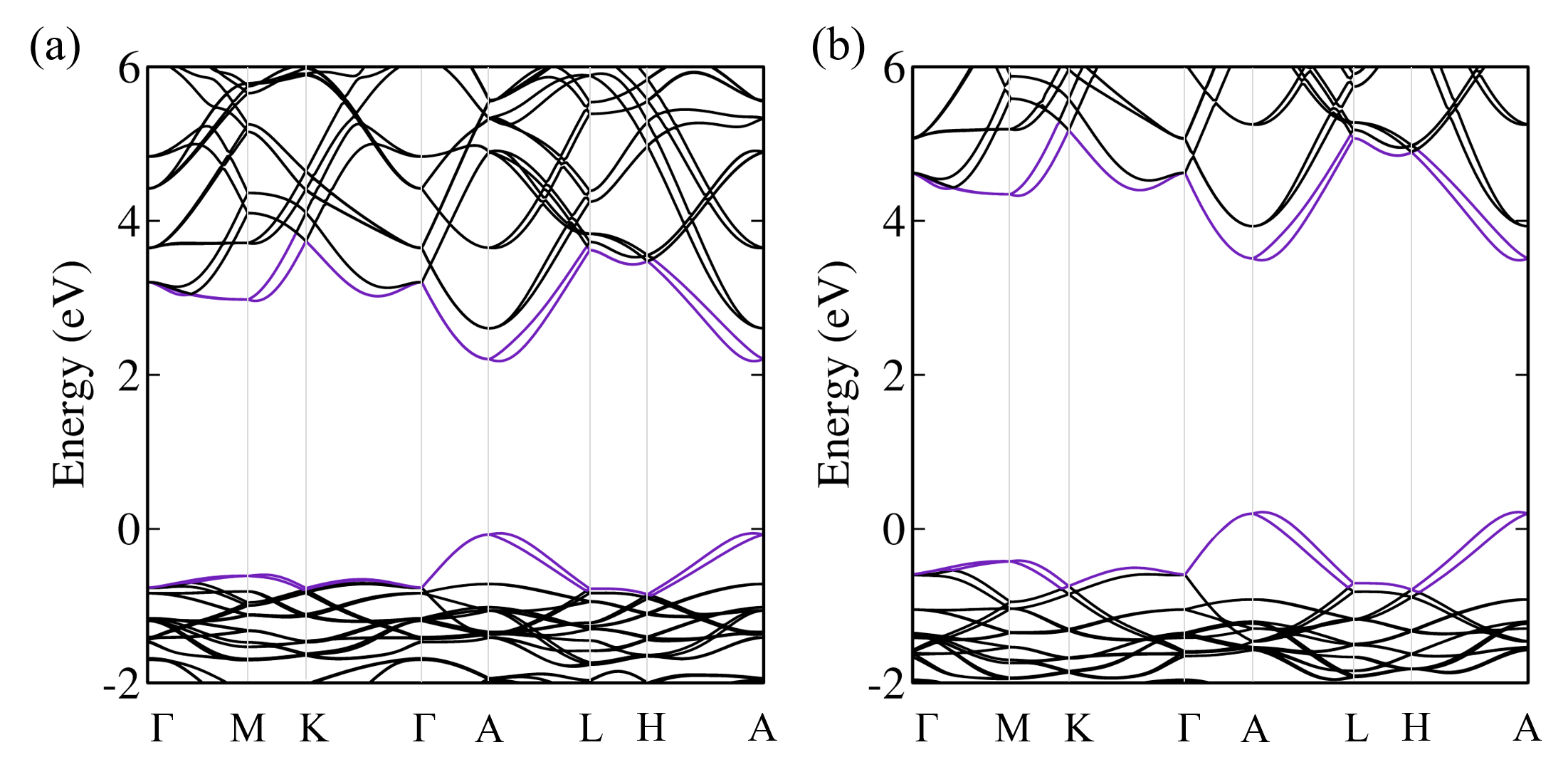}
	\caption{Band structure for $R3m$ phase of KIO$_3$ calculated using (a) PBE+SOC and (b) HSE06+SOC. The Fermi energy is set to VBM.}
	\label{fig_1}
\end{figure}

We have computed the band structure of $R3m$ phase of KIO$_3$ using semi-local Perdew-Burke-Ernzerhof (PBE) and non-local Heyd-Scuseria-Ernzerhof (HSE06) exchange-correlation ($\epsilon_{xc}$) functional combining with spin-orbit coupling (SOC).  Figure \ref{fig_1}a and \ref{fig_1}b show the band structures calculated using PBE+SOC and HSE06+SOC, respectively. A slightly indirect band gap of 2.24 and 3.27 eV is observed using PBE+SOC and HSE06+SOC, respectively, at $k$-point A. %Rashba parameters are compared using both functionals for conduction band minimum (CBm). 
Rashba spin-splitting energy ($\delta E$), offset momentum ($\delta k$) along A-H and A-L directions, Rashba coefficient ($\alpha_R$) and Dresselhaus coefficient ($\alpha_D$) at conduction band minimum (CBm) are compared using both functionals in Table~\ref{T3}. Since Rashba parameters are comparable using both functionals, we have used PBE+SOC for further calculations being more cost effective. 

\begin{table}[H]
	\begin{center}
		\caption {Rashba parameters for conduction band-splitting at \textit{k}-point A for $R3m$ phase.}
		\label{T3}
		\begin{tabular}{|p{2.50cm}|p{1.25cm}|p{1.25cm}|p{1.25cm}|p{1.20cm}|p{1.20cm}|}
			\hline
			Functional& $\delta E$ & $\delta$\textit{k}$_{\textrm{A-L}}$ & $\delta$\textit{k}$_\textrm{A-H}$ & $\alpha_R$& $\alpha_D$\\
			& (meV) &(\AA$^{-1}$)& (\AA$^{-1}$) & (eV\AA) & (eV\AA)  \\ \hline
			PBE+SOC        &  23.2  & 0.054  & 0.042& 0.77     &     0.11       \\ \hline
			HSE06+SOC        & 24.1   & 0.056   & 0.043 & 0.79     &  0.08              \\ \hline
		\end{tabular}
	\end{center}
\end{table}
\section{Two-band \textbf{\textit{k.p}} Hamiltonian} We aim at deriving the symmetry adapted two-band effective model Hamiltonian around the high symmetry $k$-point A. All those terms are included which are invariant under symmetry group operation, i.e., $O^{\dagger}H(\textbf{\textit{k}})O=H({\textbf{\textit{k}}})$. Here, \textit{O} represents the symmetry group operations belonging to the group of wave vectors and time-reversal operations. It is noteworthy that the considered \textbf{\textit{k.p}} 
Hamiltonian  includes only spin degrees of freedom and does not take into account the orbital degrees of freedom. We have incorporated all the terms of the form $k^n_{\alpha}\sigma_{\beta}$, where k$_\alpha$ and $\sigma_{\beta}$ are the crystal momenta and Pauli spin matrices, respectively, along with the free particle Hamiltonian $H_o(k)$. Since the time-reversal operator transforms the $k_\alpha$ to $-k_\alpha$ and $\sigma_{\beta}$ to -$\sigma_{\beta}$, the terms which are odd in momentum space, are only allowed to hold the time-reversal symmetry. So the general expression of two-band \textbf{\textit{k.p}} model can be written as
\begin{equation}
H(\textbf{\textit{k}})=H_o(\textbf{\textit{k}})+\sum k^n_{\alpha}\sigma_{\beta}
\end{equation}

where $\alpha$, $\beta$ = $x$, $y$, $z$ and $n$ takes only odd positive integers. In this model, for the specific case of $k$-point A, we have included upto cubic terms in crystal momentum, i.e., $k_x\sigma _x$, $k_x\sigma _y$, $k_x\sigma _z$, $k_y\sigma _x$, $k_y\sigma _y$..., $k_z^3\sigma _z$.  Point group symmetry at $k$-point A is C$_{3v}$, which can be generated by trivial identity operation (E), three fold rotation about \textit{z}-axis (C$_{3z}$=$e^{{-i\pi}/{3\sigma_z}}$) and mirror plane reflection in \textit{y-z} plane (M$_x$=$i\sigma_x$) ~\cite{link1}.  Transformation rules for $\sigma_{\beta}$ and $k_\alpha$ under C$_{3v}$ point group and time-reversal operations are summarized in Table~\ref{T1}. 
\begin{table}
	\begin{center}
	\caption {The transformations of ($\sigma_x$, $\sigma_y$, $\sigma_z$) and ($k_x$, $k_y$, $k_z$) with respect to the generators of the C$_{3v}$ point group and time-reversal operator (T). Note that the generators C$_{3z}$ and M$_x$ are enough to form the whole group of C$_{3v}$. Hence, only these generators with time-reversal T=$i\sigma_y$K operation (K is complex conjugation operator) are considered to construct the \textbf{\textit{k.p}} model for the $k$-point A.}
		\label{T1}
		\begin{tabular}{ |c|c|c|c|}
			\hline
			& C$_{3z}$=$e^{{-i\pi}/{3\sigma_z}}$ & M$_x$=$i\sigma_x$ & T=$i\sigma_y$K           \\ \hline
			$k_x$       &  -$k_x$/2+$\sqrt{3}k_y$/2  &-k$_x$ & -k$_x$        \\ \hline
			$k_y$       &  -$\sqrt{3}k_x$/2-$k_y$/2  & k$_y$ & -k$_y$      \\ \hline
			$k_z$       &  $k_z$  & $k_z$ & -k$_z$       \\ \hline
			$\sigma_x$  &  -$\sigma_x$/2+$\sqrt{3}\sigma_y$/2   & $\sigma_x$   & -$\sigma_x$   \\ \hline
			$\sigma_y$  &  -$\sqrt{3}\sigma_x$/2-$\sigma_y$/2  & -$\sigma_y$   & -$\sigma_y$   \\ \hline
			$\sigma_z$  &  $\sigma_z$  & -$\sigma_z$   & -$\sigma_z$  \\ \hline
		\end{tabular}
	\end{center}
\end{table}
Thus, the constructed Hamiltonian can be written as 
\begin{equation}
H_A(\textbf{\textit{k}})=H_o(\textbf{\textit{k}})+ H_{SO}
\label{Eq_1}
\end{equation}
where, 
\begin{equation}
H_{SO}=\alpha \sigma_y k_x + \beta \sigma_x k_y 
+ \gamma \sigma_z f(k_x,k_y)
\end{equation}
and $H_o(\textbf{\textit{k}})$ is free particle Hamiltonian. Here, $\alpha$, $\beta$ are the coefficients  of linear terms and $\gamma$ is the coefficient of cubic term in SOC Hamiltonian. The linear Rashba and Dresselhaus Hamiltonian are given by $\alpha_R$($\sigma_xk_y-\sigma_yk_x$) and $\alpha_D$($\sigma_xk_y+\sigma_yk_x$), respectively~\cite{tao2021perspectives}. Here, $\alpha_R$ and $\alpha_D$ are linear Rashba and Dresselhaus coefficients, respectively, which depend upon the properties of materials. The function $f(k_x,k_y)$, which has cubic dependence on crystal momentum, is given by
\begin{equation}
f(k_x,k_y)=(k_x^3+k_y^3)-3(k_xk_y^2+k_yk_x^2)
\label{4}
\end{equation}  Writing the Hamiltonian in matrix representation

\begin{equation}
H_A(\textbf{\textit{k}})=\begin{pmatrix}
E_0(\textbf{\textit{k}})-\gamma f &  \beta k_y-i\alpha k_x\\
\beta k_y-i\alpha k_x  & E_0(\textbf{\textit{k}})-\gamma f
\label{eq1}
\end{pmatrix}
\end{equation}
where the $E_0(\textbf{\textit{k}})=\frac{\hbar^2 k_x^2}{2m_x}+\frac{\hbar^2 k_y^2}{2m_y}$ is the energy eigenvalue of free particle Hamiltonian. On diagonalizing the Hamiltonian, i.e matrix in Eq. \ref{eq1}, gives

\begin{equation}
E(\textbf{\textit{k}})^{\pm}= E_0(\textbf{\textit{k}}) \pm \sqrt{\alpha^2 k_x^2+\beta^2 k_y^2+ \gamma^2 f^2(k_x,k_y)}  
\end{equation} and the corresponding spinor eigenfunctions are given by 

\begin{equation}
\Psi_{\textbf{\textit{k}}}^{\pm} = \frac{e^{i\textbf{\textit{k}.r}}}{\sqrt{2\pi(\rho^2_{\pm}+1 })}\begin{pmatrix}
\frac{i\alpha k_x-\beta k_y}{\gamma f(k_x,k_y)\mp E_{SO}} \\
1
\end{pmatrix}
\end{equation}
where $\rho^2_{\pm}=\frac{\alpha^2 k_x^2+\beta^2 k_y^2}{(\gamma f(k_x,k_y)\mp E_{SO})^2}$ is normalization constant, $E_{SO}$ = {$\mid E(\textbf{\textit{k}})-E_0(\textbf{\textit{k}})\mid$} is the absolute energy eigenvalue of spin-orbit coupling Hamiltonian. Spin textures can be computed using the expectation values of spin operators. Using the fact that $S_i=\frac{\sigma_i}{2}$, expectation values of $S_i$ ($s_i$=$\langle S_i \rangle$) are given by

\begin{equation}
 \{s_x,s_y,s_z\}^{\pm}= \frac{1}{2}\{{\langle \sigma_x \rangle^{\pm}},\langle \sigma_y \rangle^{\pm},\langle \sigma_z \rangle^{\pm} \}= \pm \frac{1}{E_{so}}\{\beta k_y,\alpha k_x,\gamma f(k_x,k_y) \}
 \label{8}
\end{equation}

where $\langle \sigma_i \rangle^{\pm}$= $\bra{ \Psi_{\textbf{\textit{k}}}^{\pm}} \sigma_i \ket {\Psi_{\textbf{\textit{k}}}^{\pm}}$ are the expectation values of the spin operators. Using Eq. \ref{4} and \ref{8}, we can say that the three-fold degeneracy of out of plane spin texture ($S_z$) is a consequence of cubic term $f(k_x,k_y)$. The out of plane spin component is zero, when $S_z=0$ or $f(k_x,k_y)=0$. The lines L$_1$: $k_y=-k_x$, L$_2$: $k_y=2x-\sqrt{3}x$ and L$_3$: $k_y=2x+\sqrt{3}x$ in the momentum space are the directions of out of plane spin component. Slope of the lines L$_1$, L$_2$ and L$_3$ are -1, 2-$\sqrt{3}$ and 2+$\sqrt{3}$, respectively. Angle between two lines of slopes $m_1$ and $m_2$  can be computed using $\theta=tan^{-1}\mid \frac{m_1-m_2}{1+m_1m_2} \mid$. It is straightforward to see that smaller angle between any two lines is $60\degree$, confirming the existence of three-fold degeneracy.\\

Cubic terms have insignificant role in band splitting near the high symmetry point A because near $k$-point A, $\mid k_x \mid$ and $\mid k_y \mid$ $\ll $ 1. Hence, only the linear part of Hamiltonian contributes in calculating $\alpha_R$ and $\alpha_D$. $\alpha_R$ and $\alpha_D$ can be obtained using the expressions given below by considering only the linear terms.

\begin{equation}
H_{SO}({\textbf{\textit{k}}})= H_R+H_D= \alpha \sigma_y k_x + \beta \sigma_x k_y = ({\alpha_R+\alpha_D})\sigma_y k_x+  ({\alpha_D-\alpha_R})\sigma_x k_y
\end{equation}

Thus, comparison of the coefficients gives
\begin{equation}
\alpha_R=\frac{\alpha-\beta}{2}
\end{equation}
and

\begin{equation}
\alpha_D=\frac{\alpha+\beta}{2}
\end{equation}

Note that the cubic terms play insignificant role in band splitting and must  be included for explaining the out of plane spin component as discussed earlier.

\section{Rashba parameters for selected bulk ferroelectric materials}

In the Table~\ref{T2}, we have compared the Rashba coefficients ($\alpha_R$) of some well known bulk ferroelectric materials with the bulk KIO$_3$. Space group symmetry is also included with the material. For  hafnia (HfO$_2$), $\delta E$ and $\delta k$ are not provided in the literature. All the values are reported upto respective significant figures in the references and may not be consistent with each other.  

\begin{table}[H]
	\begin{center}
		\caption {Rashba spin-splitting energy ($\delta E$), offset momentum ($\delta k$) and Rashba coefficient ($\alpha_R$) of some selected bulk ferroelectric materials.}
		\label{T2}
		\begin{tabular}{ |c|c|c|c|c|c|}
			\hline
			
			Material& Space group & $\delta E$ (meV) & $\delta k$ (\AA$^{-1}$) &  $\alpha_R$ (eV\AA)&  Reference      \\ \hline
			KIO$_3$ &$R3m$  &  $23.2$  & 0.053   & 0.77 & This work \\ \hline
			KIO$_3$&$R3c$  & 14.8   &  0.047  & 0.52 & This work  \\ \hline
			BiAlO$_3$&$R3c$       &  7.34  &0.04 & 0.39 &  \cite{da2016rashba}     \\ \hline
			BiAlO$_3$ (along Z-R)&$P4mm$       &  9.40  & 0.03 & 0.74 & \cite{da2016rashba} \\ \hline
			BiAlO$_3$ (along A-Z)&$P4mm$       &  8.62  & 0.03 & 0.65& \cite{da2016rashba}  \\ \hline
			LaWN$_3$&$Pna2_1$      &  2.20  & 0.014  & 0.31  & \cite{bandyopadhyay2020origin} \\ \hline
			LaWN$_3$&$R3c$  &  3.49  & 0.051  & 0.18 & \cite{bandyopadhyay2020origin} \\ \hline
			BiInO$_3$&$Pna2_1$  &  130 & 0.19  & 1.91 & \cite{tao2018persistent} \\ \hline
			PbTiO$_3$&$P4mm$ &  5.45  & 0.50   & 0.51 & \cite{arras2019rashba}\\ \hline
			HfO$_2$&$Pca2_1$  &  -  & -  & 0.06 & \cite{tao2017reversible} \\ \hline
			KMgSb&$P6_3mc$  &  10  & 0.024   & 0.83 & \cite{narayan2015class} \\ \hline
			LiZnSb&$P6_3mc$ &  21  & 0.023   & 1.82 & \cite{narayan2015class} \\ \hline
			
			$\beta$-(MA)PbI$_3$&$P4mm$ &  12 & 0.015  & 1.5 & \cite{kim2014switchable} \\ \hline
			$\beta$-(MA)SnI$_3$&$P4mm$ &  11  & 0.011   & 1.9 & \cite{kim2014switchable} \\ \hline
			GeTe&$R3m$  &  227 & 0.09   & 4.8 & \cite{di2013electric} \\ \hline
			
		\end{tabular}
	\end{center}
\end{table}

\bibliography{SIref}

\begin{figure}[h!]
\includegraphics[width=0.8\columnwidth,clip]{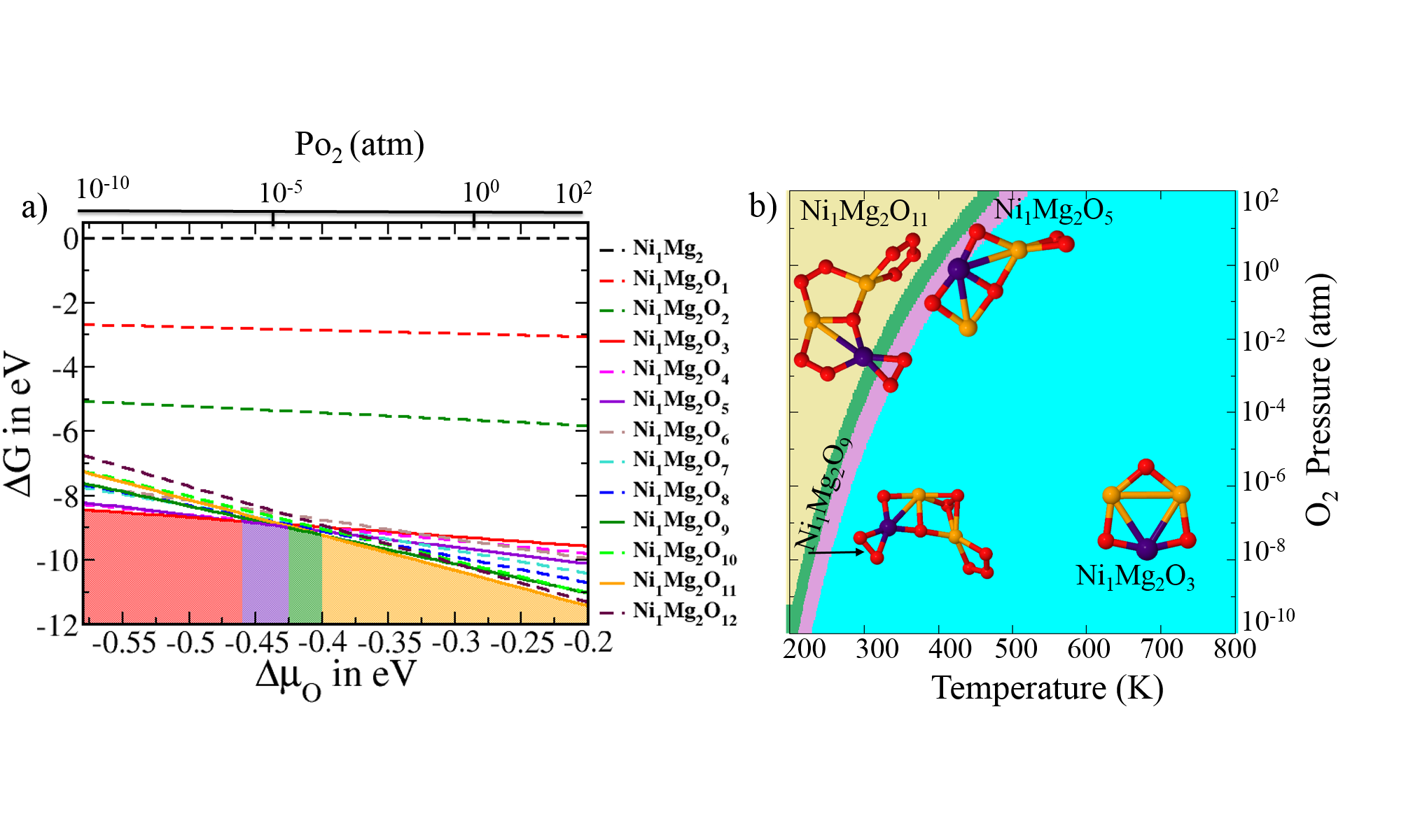}
\caption{In a), $\Delta G$ as a function of chemical potential of oxygen. Total energy of geometries are calculated using PBE+vdW. Colored areas are for the guideline to identify which phase is the most stable phase for different range of pressure. In b), 3D phase diagram for Ni$_1$Mg$_2$O$_x$ clusters}
\label{fig:SIpic1}
\end{figure}

\title{Small ternary $TM_x$Mg$_y$O$_z$ clusters ($TM$ = Cr, Ni, Fe, Co; $x+y \leq 3$) at realistic conditions: Unraveling stability and electronic structure from first-principles}
\author{Shikha Saini, Debalaya Sarker, Pooja Basera and Saswata Bhattacharya\footnote{saswata@physics.iitd.ac.in}} 
\affiliation{Ab initio Research Group, Department of Physics, Indian Institute of Technology Delhi, Hauz Khas 110016, New Delhi, India}
\date{\today}
\pacs{}
\keywords{CH$_3$NH$_3$PbI$_3$, CH$_3$NH$_3$SnI$_3$, defects, DFT, vacancy, carrier concentration, free energy}
\maketitle
%%%%%%%%%%%%%%%%%
\begin{center}
{\Large \bf Supplemental Material}\\ 
\end{center}
\vspace*{12pt}
\begin{enumerate}[\bf I.]

\item Effect of functionals on phase diagrams.

\end{enumerate}
%\clearpage
%\newpage
%%%%%%%%%%%%%%
\noindent {\bf \Large I. Effect of functionals on phase diagrams\\}
In the FIG. \ref{fig:SIpic1}a) below, we have plotted the free energy of formation $\Delta G$ of globally optimized clusters (Ni$_1$Mg$_2$O$_x$) with varying oxygen chemical potential ($\Delta {{\mu }_{{\rm O}}}$) at a finite temperature (T=300K). In order to evaluate $\Delta G$, we have calculated free energy of the clusters Ni$_1$Mg$_2$O$_x$ (for x=1,2,3...) that is approximated by their $E\textsuperscript{DFT}$ (DFT total energy) and $F\textsuperscript{vibrational}$ (see Eq. 1 and 2). Here, the electronic energies are calculated by using PBE+vdW functional. 
\begin{equation}
	\begin{split}
		\Delta G_f\left(T,p_{O_2}\right) = F_{Me_xMg_yO_z}(T) - F_{Me_xMg_y}(T) - z\Delta\mu_O\left(T,p_{O_2}\right)
		\label{eq2}
	\end{split}
\end{equation}
where,
\begin{equation}
	\begin{split}
		F(T) =  F\textsuperscript{vibrational}(T) + E \textsuperscript{DFT}
		\label{eq3}
	\end{split}
\end{equation}
The chemical potential of oxygen is calculated by using Eq. 3. Where, $\nu_{OO}$ is the O-O stretching frequency and $\frac{h\nu_{OO}}{2}$ is the zero point energy of the oxygen molecule.
\begin{equation}
	\begin{split}
		\Delta \mu_O \left(T,p_{O_2}\right) & =\frac{1}{2}\left(\mu_{O_2} \left(T,p_{O_2}\right) - E^\textsuperscript{DFT}_{O_2} - \dfrac{h\nu_{OO}}{2}\right)
		\label{eq4}
	\end{split}
\end{equation}
Where,
\begin{equation}
	\begin{split}
		\mu_{O_2}\left(T,p_{O_2}\right) & = -k_BT ln \left[\left(\frac{2\pi m}{h^2}\right)^\frac{3}{2} (k_BT)^\frac{5}{2}\right] + k_BT ln p_{O_2} - k_BT ln \left(\frac{8 \pi^2 I_Ak_BT}{h^2} \right)\\& + \frac{h\nu_{OO}}{2} + k_BT ln\left[ 1 - \exp \left(-\frac{h\nu_{OO}}{k_BT}\right)\right]+ E \textsuperscript{DFT} -k_BT ln \mathcal{M} 
		+k_BT ln \sigma 
		\label{eq5}
	\end{split}
\end{equation}

\begin{figure}[h!]
	\includegraphics[width=0.8\columnwidth,clip]{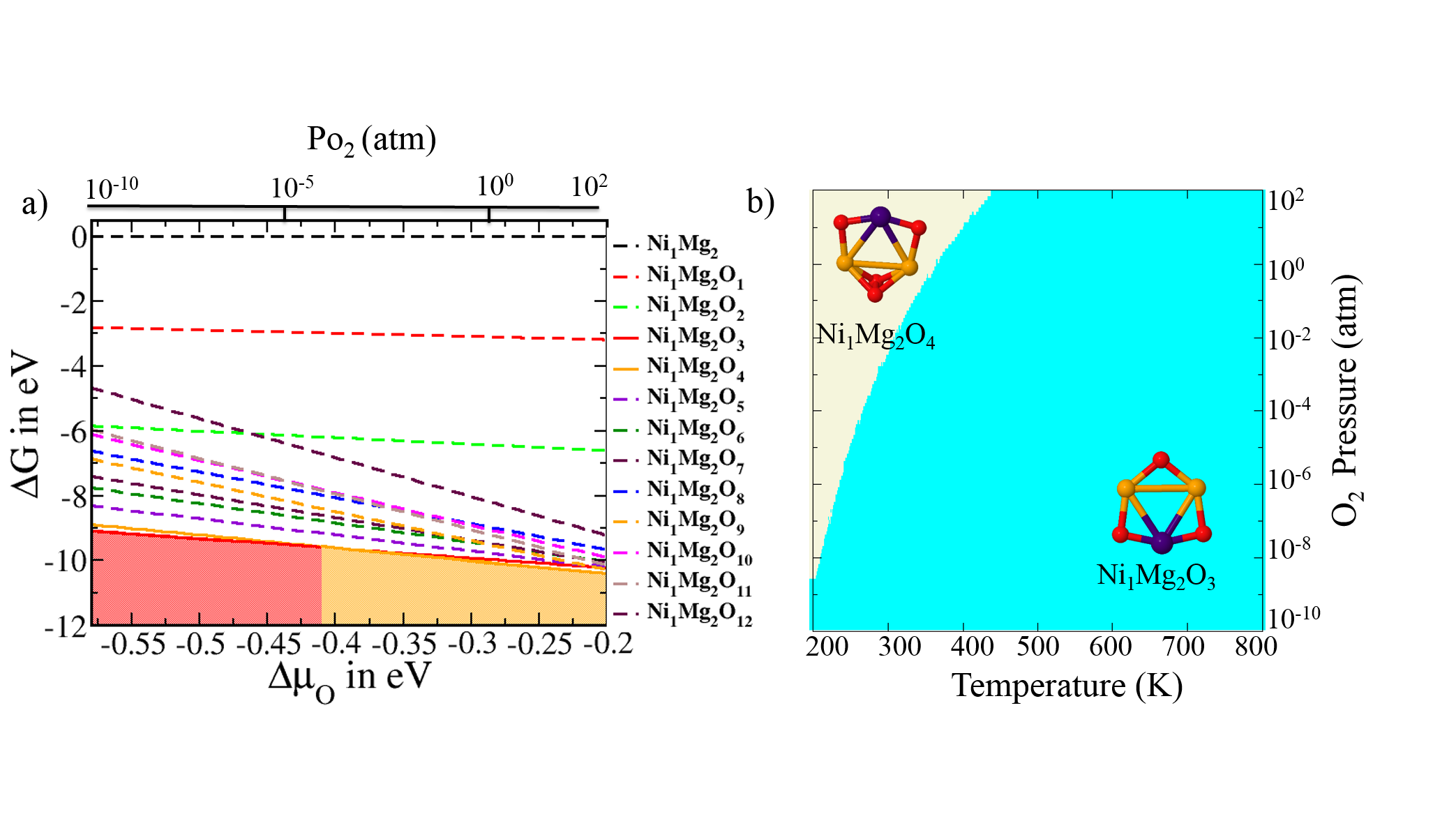}
	\caption{2D (a) and 3D (b) phase diagram for same set of clusters. Total energy of geometries are calculated using HSE06+vdW.}
	\label{fig:SIpic2}
\end{figure}

 In the above equation, $m$ is the mass, $I_A$ is the moment of inertia, $\mathcal{M}$ is the spin multiplicity, and $\sigma$ is the symmetry number for oxygen molecule. Pressure dependency in the free energy of formation $\Delta G$ comes in $\Delta\mu_O$ expression only. Therefore, the pressure axes are calculated according to the relation between $\mu_{O_2}$, with ${{p}_{{{{\rm O}}_{2}}}}$ as shown in Eq. 3 and 4. This variation of pressure in chemical potential ($\Delta {{\mu }_{{\rm O}}}$) is shown on top x-axis in the FIG. \ref{fig:SIpic1}a). The structures and compositions, which exhibit the lowest free energy of formation $\Delta G$ in the experimentally accessible (T, ${{p}_{{{{\rm O}}_{2}}}}$) region, is the most stable one under that (T, ${{p}_{{{{\rm O}}_{2}}}}$) region. In lower range of pressure Ni$_1$Mg$_2$O$_3$ is the most stable one, as we increase the pressure Ni$_1$Mg$_2$O$_5$, Ni$_1$Mg$_2$O$_9$ and, finally Ni$_1$Mg$_2$O$_{11}$ compositions are more preferable as shown in FIG. \ref{fig:SIpic1}a). Lowest lying lines are corresponding to the most stable phases as shown in FIG. \ref{fig:SIpic1}a). In FIG. \ref{fig:SIpic1}b), the dependence of temperature and pressure are combined in a 3D phase diagram. In 3D phase diagram, we could able only to concentrate on the stable phase with varying temperature and pressure simultaneously within single figure. In FIG. \ref{fig:SIpic1}b), there is a color corresponding to a particular composition, that indicates which phase is the most stable phase at which T and ${{p}_{{{{\rm O}}_{2}}}}$. In FIG. 1b) at high temperature and all range of pressure Ni$_1$Mg$_2$O$_3$ is the most stable phase, and at lower T and ${{p}_{{{{\rm O}}_{2}}}}$  Ni$_1$Mg$_2$O$_5$, Ni$_1$Mg$_2$O$_9$ are the most favorable. Further, as we increase the pressure at lower range od temperature Ni$_1$Mg$_2$O$_{11}$ phase is more preferable.
Same as in FIG. \ref{fig:SIpic1}, we have plotted the 2D and 3D phase diagrams by using more advanced hybrid functional (HSE06+vdW). Here, we have calculated the electronic energies of Ni$_1$Mg$_2$O$_x$ clusters by using HSE06+vdW functional. From the comparison of FIG. \ref{fig:SIpic1} and FIG. \ref{fig:SIpic2}, it can be clearly seen that there are huge difference in the phase diagrams with different functionals. In FIG. \ref{fig:SIpic2}a), at lower pressure range Ni$_1$Mg$_2$O$_3$ is the most stable phase at 300K. As we increase the pressure Ni$_1$Mg$_2$O$_4$ is the favorable phase. From FIG. \ref{fig:SIpic2}b) we can see the most stable phases at all range of temperature with varying the pressure. At lower range of pressure and all temperature Ni$_1$Mg$_2$O$_3$ is most stable one and as we increase the pressure at lower range of temperature Ni$_1$Mg$_2$O$_4$ is the favorable phase.  
\bibliography{Bibliography}{}